\documentclass[amssymb,amsmath,prl,twocolumn,floatfix,superscriptaddress]{revtex4}
\usepackage{graphicx}
\usepackage{lastpage}
\usepackage{amssymb}

\begin{document}

\title{Quantum phase transitions of light.}

\author{Andrew D.~Greentree}

\email{andrew.greentree@ph.unimbelb.edu.au}

\affiliation{Centre for Quantum Computer Technology, School of Physics, The University
of Melbourne, Victoria 3010, Australia.}

\author{Charles Tahan}

\affiliation{Cavendish Laboratory, University of Cambridge, JJ Thomson Ave, Cambridge
CB3 0HE, United Kingdom}

\affiliation{Centre for Quantum Computer Technology, School of Physics, The University
of Melbourne, Victoria 3010, Australia.}

\author{Jared H. Cole}

\affiliation{Centre for Quantum Computer Technology, School of Physics, The University
of Melbourne, Victoria 3010, Australia.}

\author{L. C. L. Hollenberg}

\affiliation{Centre for Quantum Computer Technology, School of Physics, The University
of Melbourne, Victoria 3010, Australia.}

\date{\today{}}

\maketitle \textbf{Recently, condensed matter and atomic experiments
have reached a length-scale and temperature regime where new quantum
collective phenomena emerge. Finding such physics in systems of
photons, however, is problematic, as photons typically do not
interact with each other and can be created or destroyed at will.
Here, we introduce a physical system of photons that exhibits
strongly correlated dynamics on a meso-scale. By adding photons to a
two-dimensional array of coupled optical cavities each containing a
single two-level atom in the photon-blockade regime, we form dressed
states, or polaritons, that are both long-lived and strongly
interacting. Our zero temperature results predict that this photonic
system will undergo a characteristic Mott insulator (excitations
localised on each site) to superfluid (excitations delocalised
across the lattice) quantum phase transition. Each cavity's
impressive photon out-coupling potential may lead to actual devices
based on these quantum many-body effects, as well as observable,
tunable quantum simulators.}

The Jaynes-Cummings \cite{bib:JC} model is arguably the most
important model for understanding light-matter interactions. It
describes the interaction of a single, quasi-resonant optical cavity
field with a two-level atom. The coupling between the atom and the
photons leads to optical nonlinearities and an effective
photon-photon repulsion. Perhaps the most extreme demonstration of
this photonic repulsion is photon blockade, demonstrated recently by
Birnbaum \textit{et al.} \cite{bib:BirnbaumNature2005}, where
photonic repulsion prevents more than one photon from being in the
cavity at any one time. Photon blockade was initially theoretically
described with a four-state system \cite{bib:ImamogluPRL1997}, with
multiplication of the weak Kerr nonlinearity effected by placing a
large number of atoms within each cavity. However, it was quickly
realised that the photonic blockade mechanism does not persist in
the limit of many atoms \cite{bib:GrangierPRL1998}, rapidly
degrading as the number of atoms per cavity is increased
\cite{bib:GreentreeJOPB2000}. Later Rebic \textit{et al.} showed
that the nonlinear interaction afforded by placing a single
two-level atom inside a cavity would suffice for realising photon
blockade \cite{bib:RebicPRA2002}. This observation was highly
significant as it allowed the full weight of the Jaynes-Cummings
model to be used to attack and understand this problem.

\begin{figure}[tb!]
\includegraphics[width=0.8\columnwidth,clip]{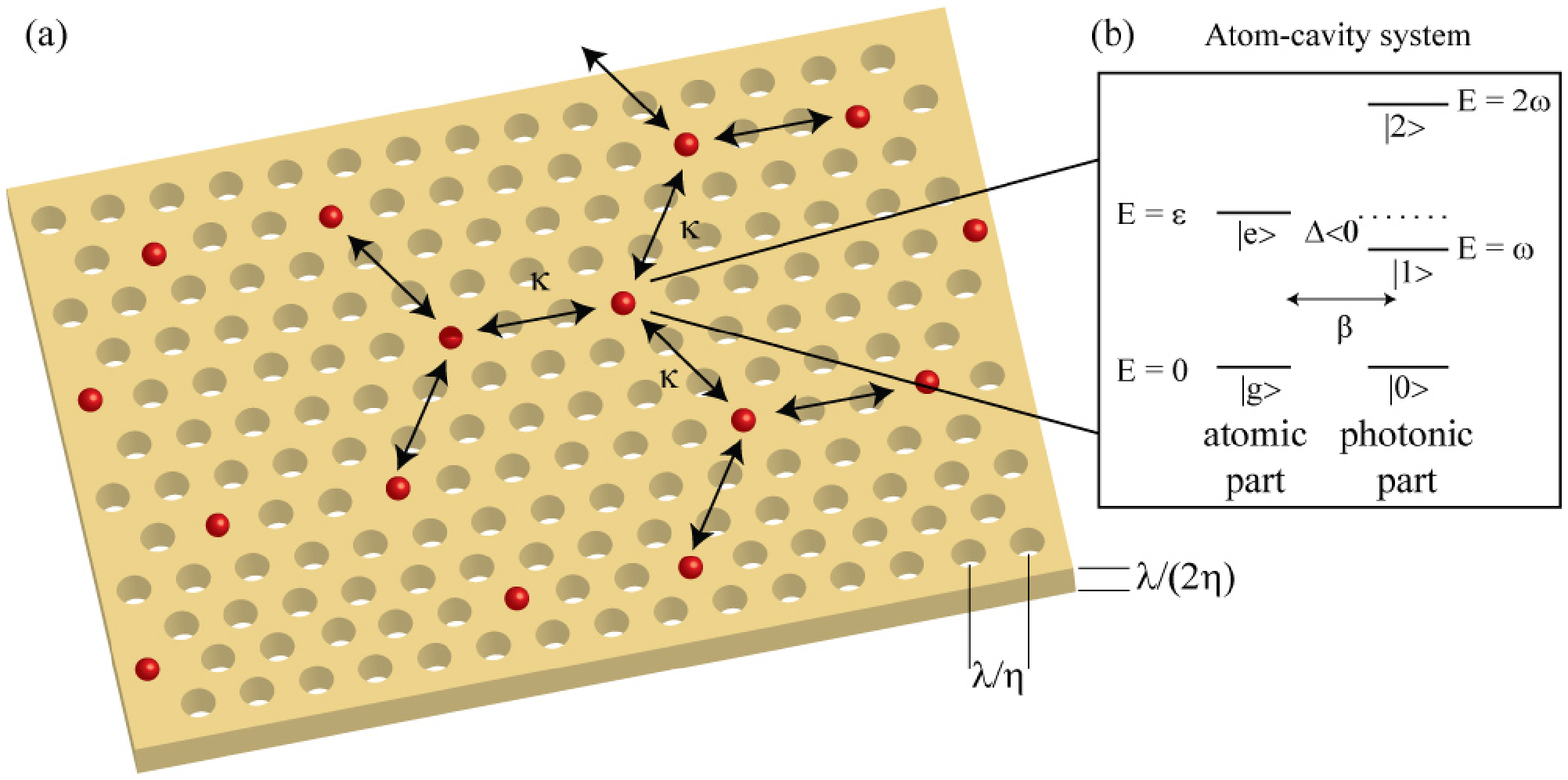}
\caption{\label{fig:Fig1} (a) Schematic showing a two-dimensional
array of photonic bandgap cavities, with each cavity containing a
single two-level atom (spheres).  The photonic bandgap is described
by a periodic perturbations of the dielectric medium, in this case
represented by the regions drilled through a thin membrane, for
example via focussed ion beam.  The spacing between the holes will
be of order $\lambda/\eta$ where $\lambda$ is the optical wavelength
and $\eta$ the refractive index.  The membrane should be of order
$\lambda/(2\eta)$ thick to confine light in the plane. Cavities are
defined by defects in the periodic modulation, i.e. where a rod has
\emph{not} been drilled. By introducing a periodic array of defects
in the photonic bandgap structure we can realise a lattice of
photonic bandgap cavities, i.e. a photonic bandgap cavity
superlattice. Photons can hop from one cavity to either of the three
nearest neighbours with a frequency (photon hopping rate) $\kappa$,
illustrated for some of the cavities by the double headed arrows. To
effect photon-photon repulsion, we insert a single two-state system
into each cavity, represented here by the red spheres.  If the
membrane were fabricated in diamond, the two-state systems could be
ion implanted nitrogen-vacancy (NV) centres. In this lattice we have
chosen for each cavity to have three nearest neighbours, so that the
regime where photonic repulsion dominates over hopping is easier to
achieve. (b) Schematic showing some of the pertinent energy scales
within each atom-cavity system.}
\end{figure}

To create an atom-photon system whose dynamics mirror those
traditionally associated with strongly interacting condensed matter
systems, we consider a two-dimensional array of photonic bandgap
cavities. Each cavity contains a single two-level atom,
quasi-resonant with the cavity mode. Evanescent coupling between the
cavities due to their proximity allows inter-cavity photon hopping.
This configuration is depicted schematically in
Fig.~\ref{fig:Fig1}(a), where we have explicitly chosen three
nearest neighbours per cavity (coordination number $z=3$), for
reasons explained below. Because we are considering small cavities,
with volumes of order $\lambda^{3}$ where $\lambda$ is the
wavelength of the light, there will be strong atom-photon couplings
that will dominate over the spontaneous emission.
Fig.~\ref{fig:Fig1}(b) shows the pertinent energy scales within one
of the cavities. In the classical limit without two-level atoms, an
array of coupled photonic bandgap cavities has been described for
novel waveguiding applications
\cite{bib:LangJQE1988,bib:OzbayJQE2002}, and in the quantum regime a
two cavity arrangement has been proposed as a Q-switch
\cite{bib:GreentreePRA2006}.  We consider here for the first time,
the rich dynamics of a two dimensional lattice of \emph{quantum}
cavities.

Interest in interacting boson systems grew out of work on the
metal-insulator transition in Fermi systems \cite{Lee}.  The seminal
paper of Fisher \emph{et al.} \cite{Fisher} employed an infinite
range mean-field theory to qualitatively describe the
superfluid-insulator phase transitions of a bosonic gas in random
and periodic potentials, with and without disorder. Recently, it has
been shown that this Bose-Hubbard model can be realized in cold atom
optical lattices \cite{Jaksch}. The Bose-Hubbard Hamiltonian seeks
to describe the many-body dynamics of strongly interacting
bosons,
\begin{eqnarray}
\mathcal{H}^{BH} = \sum_{i,j}t_{ij}a_{j}^{\dag}a_{i} +
U\sum_{i}a_{i}^{\dag}a_{i}^{\dag}a_{i}a_{i},\label{eq:BHHam}
\end{eqnarray}
where $i,j$ range over all sites in the lattice (fixed in our case),
$t_{ij}$ is the hopping energy of bosons between sites $i$ and $j$,
$U$ is the on-site repulsion between particles, and $a_{i}$ is the
bosonic annihilation operator on site $i$. In our photonic-atom
superlattice, the Jaynes-Cummings interaction provides the effective
on-site repulsion, and the hopping term is represented by the
evanescent coupling between the cavities. However, our physical
system is fundamentally different from the Bose-Hubbard case: in
particular, the effective $U$ is not constant in the Jaynes-Cummings
system, but decreases as the number of photons in each cavity goes
up; and the conserved particles here are not pure bosons, but
dressed photons or polaritons, which are a mixture of the spinor
atom and photons in each cavity. Thus, the hopping term corresponds
to the overlap of these extended composite photon-atom states, which
are the conserved particles in the model, \emph{not} the photons.

To motivate the search for Hubbard model type interactions within a
superlattice of photonic bandgap cavities, we first discuss the
Hamiltonian for a single two-level atom in a quasi-resonant cavity,
\begin{eqnarray}
\mathcal{H}^{JC} =
\epsilon\sigma_{+}\sigma_{-}+\omega a^{\dag}a +
\beta\left(\sigma_{+}a+\sigma_{-}a^{\dag}\right), \label{eq:JCHam}
\end{eqnarray}
where $\sigma_{+}$ and $\sigma_{-}$ ($a$, $a^{\dag}$) correspond to
the atomic (photonic) raising and lowering operators, respectively.
The transition energy of the atomic system is $\epsilon$, the cavity
resonance is $\omega$ and the cavity mediated atom-photon coupling
is $\beta$, which is implicitly assumed to be real for our purposes.
Defining the atomic states as $|g\rangle$ and $|e\rangle$ for ground
and excited state respectively, we introduce the detuning,
$\Delta\equiv\omega-\epsilon$. The eigenstates of Eq.~\ref{eq:JCHam}
are the dressed states \cite{bib:CohenTannoudji1978}, which we label
as $|\pm,n\rangle$ defined in the Methods section, where $n$ is the
number of excitations in the cavity. The ground state for the
Jaynes-Cummings Hamiltonian is qualitatively different from the
other dressed states, implying a non-trivial form of the raising
operator \cite{bib:Hussin2005}. This constitutes a further departure
from usual Hubbard-like condensed-matter models, where the raising
operator is not dependent on the number of excitations.

\begin{figure}[tb!]
\includegraphics[width = 0.8\columnwidth,clip]{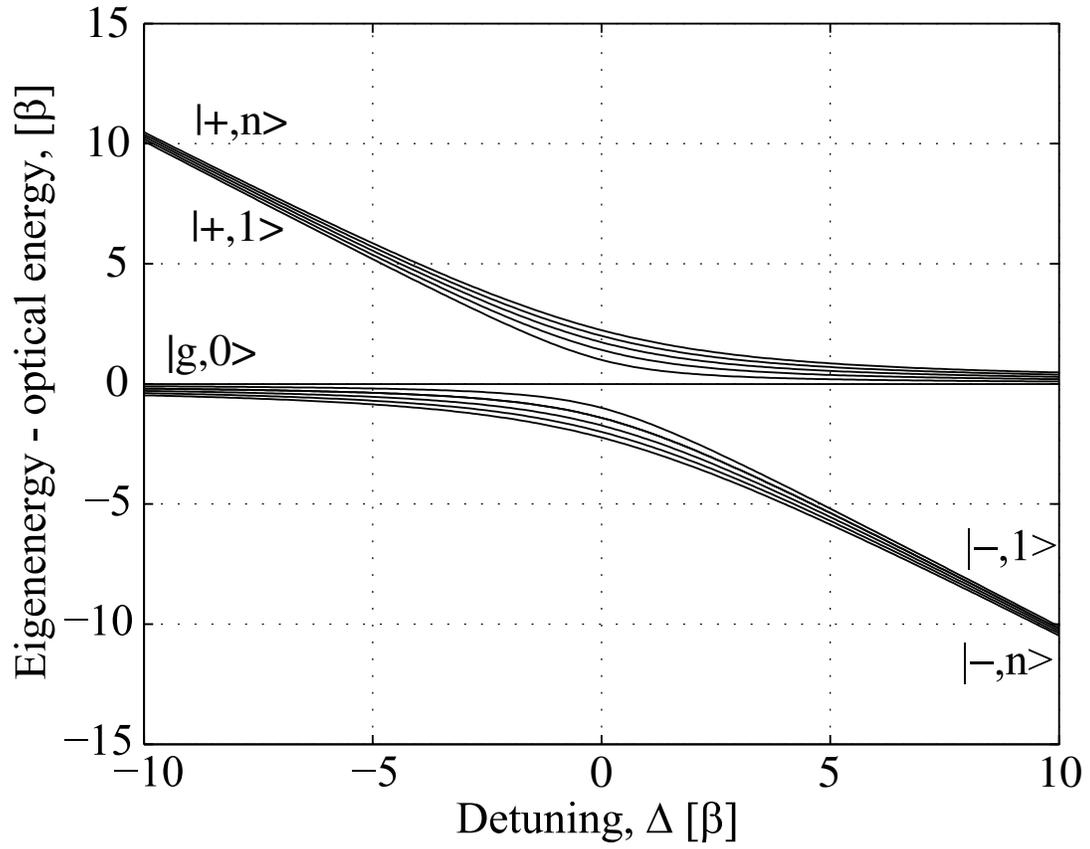}
\caption{\label{fig:Fig2} Eigenspectrum for a single atom in a high
Q cavity, as a function of the atom-cavity detuning, centered around
0, i.e. we plot $E - n\omega$. The eigenspectrum splits naturally
into two branches, corresponding to the dressed states,
$|+,n\rangle$ (upper branch) and $|-,n\rangle$ (lower branch).  Note
that the ground state $|g,0\rangle$, naturally appears to be a
member of each branch, introducing a significant departure from
usual Bose-Hubbard style dynamics within each branch. Excluding
$|g,0\rangle$, the branches anti-cross at $\Delta = 0$, with the
splitting increasing with increasing excitation number, $n$, given
by the usual Rabi frequency, $\beta\sqrt{n}$.}
\end{figure}

In Fig.~\ref{fig:Fig2} we show some of the eigenvalues of the
Jaynes-Cummings system with the photon energy subtracted for ease of
comparison, i.e.~we are plotting $E_{|\pm,n\rangle}-n\omega$. The
eigenenergies are well known and are
$E_{|\pm,n\rangle}=n\omega\pm\chi(n)-\Delta/2$ where
$\chi(n)=\sqrt{n\beta^{2}+\Delta^{2}/4}$ is the $n$ photon
generalised Rabi frequency. Displaying the eigenvalues in this way
immediately allows us to connect the Jaynes-Cummings Hamiltonian
with the on-site repulsion in the Hubbard Hamiltonian. The on-site
repulsion is evinced by the increasing energy separation with $n$.

The Hamiltonian for our extended Hubbard-like system is given by a
combination of the Jaynes-Cummings Hamiltonian with photon hopping
between cavities and the chemical potential term,
\begin{eqnarray}
\mathcal{H} = \sum_{i}\mathcal{H}_{i}^{JC} + \sum_{\langle
i,j\rangle}\kappa_{ij}\left(a_{i}^{\dagger}a_{j}+a_{j}^{\dagger}a_{i}\right)-\sum_{i}\mu_i
N_{i},
\end{eqnarray}
where the inter-cavity hopping occurs with frequency
$\kappa_{ij}=\kappa$ for nearest neighbours, and $\kappa=0$
otherwise, $N_{i}$ is the total number of atomic and photonic
excitations (the conserved quantity in our system) and $\mu_{i}$ is
the chemical potential at site $i$ in the grand canonical ensemble.
For our proof of concept calculation we assume zero disorder and
$\mu_{i}=\mu$ for all sites. To most effectively explore the
important regime where on-site repulsion dominates over hopping, we
consider a lattice with as few nearest neighbours as possible to
achieve a two-dimensional network, that is three nearest neighbours.
Altering the number of nearest neighbours does not qualitatively
affect our results.

To gain insight over the properties of the full Hamiltonian, we will
employ a mean-field approximation. Mean-field theories
\cite{Oosten1,Oosten2,Sheshadri} give good qualitative and
quantitative descriptions of these systems, comparing well to Monte
Carlo simulations \cite{Krauth,Ceperley}. These mean-field
approaches have also been extended to dipolar bosons \cite{Xie} and
boson-fermion atomic mixtures \cite{Lewenstein,Fehrmann}, as well as
to the theory of exciton and exciton-polariton (electron-hole pair
plus photon) condensation \cite{bib:Littlewood}. We introduce a
super-fluid order parameter, $\psi=\langle a_{i}\rangle$, and employ
the decoupling approximation $a_{i}^{\dagger}a_{j} = \langle
a_{i}^{\dagger}\rangle a_{j}+\langle a_{j}\rangle
a_{i}^{\dagger}-\langle a_{i}^{\dagger}\rangle\langle a_{j}\rangle.$
The resulting mean-field Hamiltonian can be written as a sum over
single sites,
\begin{eqnarray}
\mathcal{H}^{MF} & = & \sum_{i}\left\{ \mathcal{H}_{i}^{JC} - z\kappa\psi\left(a_{i}^{\dagger}+a_{i}\right)+z\kappa\left|\psi\right|^{2}\right.\nonumber \\
 &  & \left.-\mu\left(a_{i}^{\dagger}a_{i}+\sigma_{i}^{+}\sigma_{i}^{-}\right)\right\} ,
\label{eq:MF}
\end{eqnarray}
where $z=3$ is the number of nearest neighbours. To obtain the
system's zero temperature properties, we use the procedure of
Refs.~\cite{Sheshadri,Oosten2} which is outlined in the Methods
section. When $\psi=0$ we have a Mott phase, characterised by a
fixed number of excitations per site with no fluctuations, and
$\psi\neq0$ indicates a superfluid phase. The boundary between the
$\psi=0$ and $\psi\neq0$ phases denotes where a quantum phase
transition in this system will occur: \emph{a quantum phase
transition of light}. In general, we expect that when photon-photon
coupling (on-site repulsion) dominates over hopping, the system
should be in a Mott phase, and when the converse is true, the system
will be in a superfluid phase.

\begin{figure}[tb!]
\includegraphics[width = 0.8\columnwidth,clip]{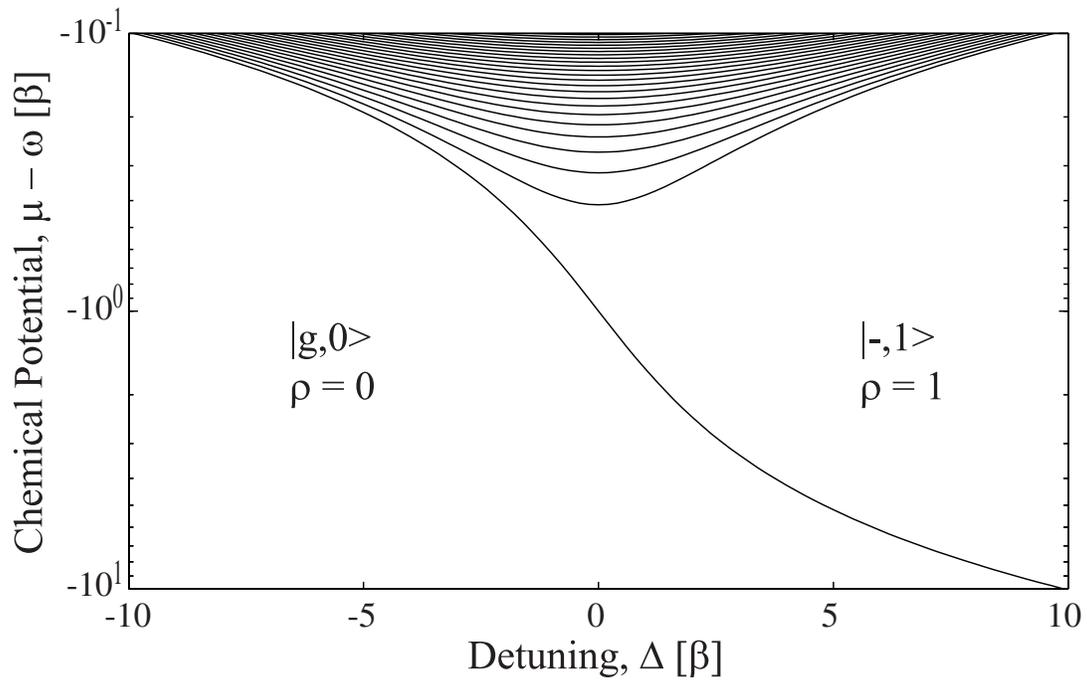}
\caption{\label{fig:Stable} Boundaries between Mott lobes in the
limit of low tunneling (small $\kappa$) as a function of $\mu$ and
$\Delta$. The two lowest stable regions correspond to $|g,0\rangle$
on the left, and $|-,1\rangle$ on the right.  The boundary between
these two domains follows the line $\mu = E_{|-,1\rangle}$. Higher
order states can be seen in the central region, defined by the
critical chemical potentials $\mu_c(n)$ and correspond to
$|-,2\rangle$, $|-,3\rangle$ etc.}
\end{figure}

\begin{figure}[tb!]
\includegraphics[width = 0.8\columnwidth,clip]{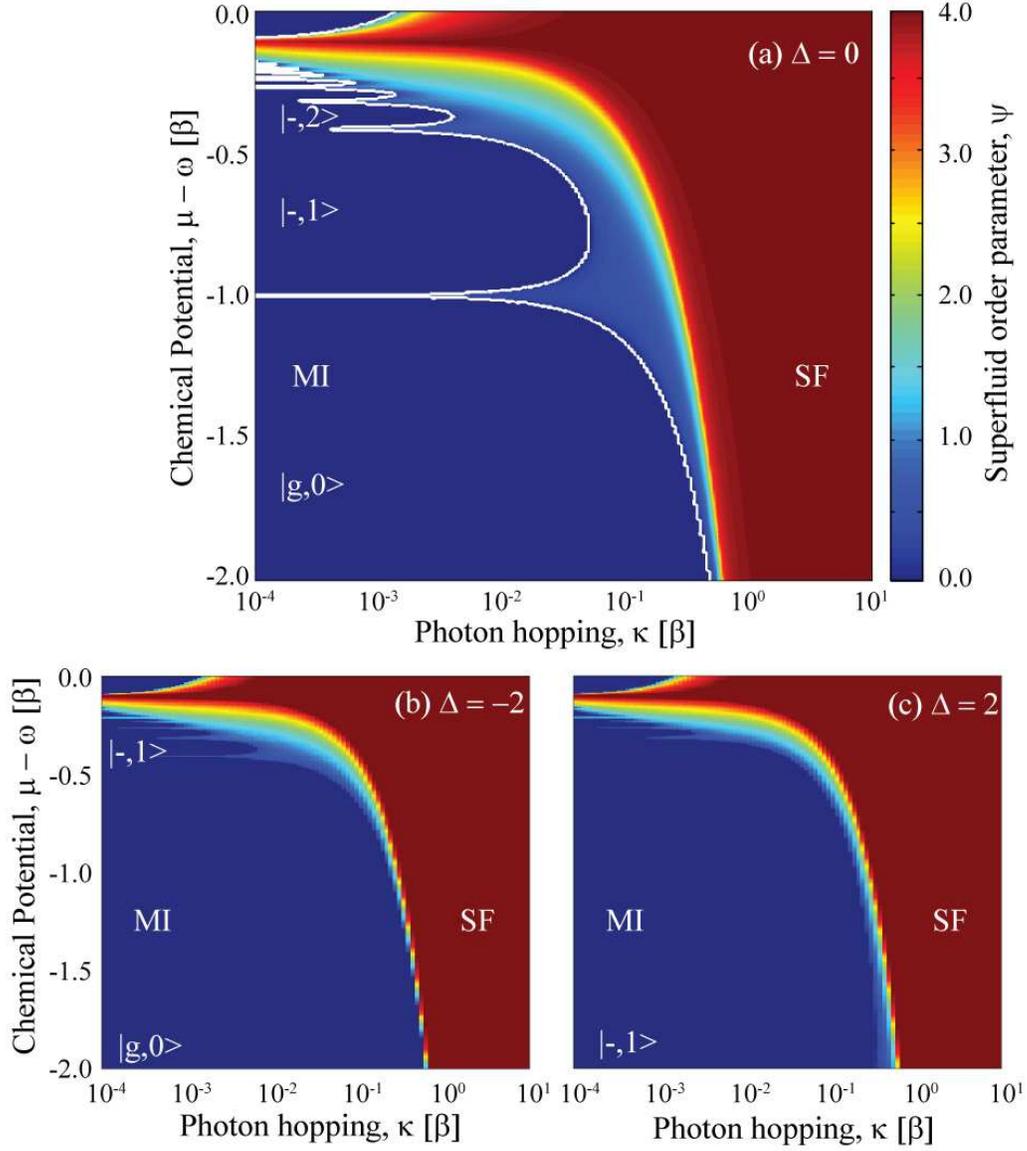}
\caption{\label{fig:Fig3} Slices showing the superfluid order
parameter, $\psi$, as a function of the photon hopping frequency,
$\kappa$, and the chemical potential, $\mu$, for (a) detuning,
$\Delta = 0$, (b) $\Delta = -2$, and (c) $\Delta = 2$. The diagrams
show Mott insulator lobes, indicated by the regions of $\psi=0$,
where the lowest have been shown. Dominating the left hand edge
(where photonic repulsion dominates over hopping) is the Mott
insulator phase (denoted MI), and the superfluid phase is found on
the right hand edge (denoted SF). The white contour in (a)
corresponds to the region where $\psi$ becomes nonzero, delineating
the quantum phase transition.}
\end{figure}

We can determine the extent of the Mott lobes in the limit of very
small hopping from inspection of Eqs.~\ref{eq:JCHam} and
\ref{eq:MF}. First note that with our definition,
$E_{|-,n\rangle}<E_{|+,n\rangle}$, we only need consider the
negative branch for the purposes of determining the ground state.
Furthermore, a change in the total number of excitations per site
will occur when $E_{|-,n+1\rangle}-\mu(n+1)=E_{|-,n\rangle}-\mu n$.
We can determine the critical chemical potential, $\mu_c(n)$, where
the system will change from $n$ to $n+1$ excitations per site as
\begin{eqnarray}
\mu_{c}(n)=\omega-\left[\chi(n+1)-\chi(n)\right]. \label{eq:mucrit}
\end{eqnarray}
These boundary chemical potentials are shown in
Fig.~\ref{fig:Stable} as a function of detuning in the limit of very
small $\kappa$, and these also account for the asymptotes in
Fig.~\ref{fig:Fig3}. In Fig.~\ref{fig:Fig3} we show the complete
phase diagram of the mean field solution. Because of the large
parameter space, there are several cases that we consider. Note that
our results are only shown for $\mu<\omega$ as in the limit
$\mu\gtrsim\omega$ the minimisation method does not converge
[$\mu_{c}(\infty)=\omega$], although in this limit, the quantum
ground state should correspond to a coherent state of excitations
(superfluid phase).

The dynamics illustrated in Fig.~\ref{fig:Fig3} are extremely rich.
Hubbard-like dynamics can be seen in the three sub-figures of
Fig.~\ref{fig:Fig3}, which show $\psi$ as a function of $\kappa$ and
$\mu$ for three different detunings. In this plot, $\psi=0$
corresponds to stable Mott lobes, with the number of excitations
increasing with $\mu$. The regions to the right correspond to
$\mu\neq0$, and in these regions the system will be found in a
superfluid phase, i.e. the stable ground state at each site
corresponds to a coherent state of excitations over the
$|-,n\rangle$ branch. The size of the Mott lobes varies with
$\Delta$, with the largest Mott lobes found on resonance.

\begin{figure}[tb!]
\includegraphics[width = 0.8\columnwidth,clip]{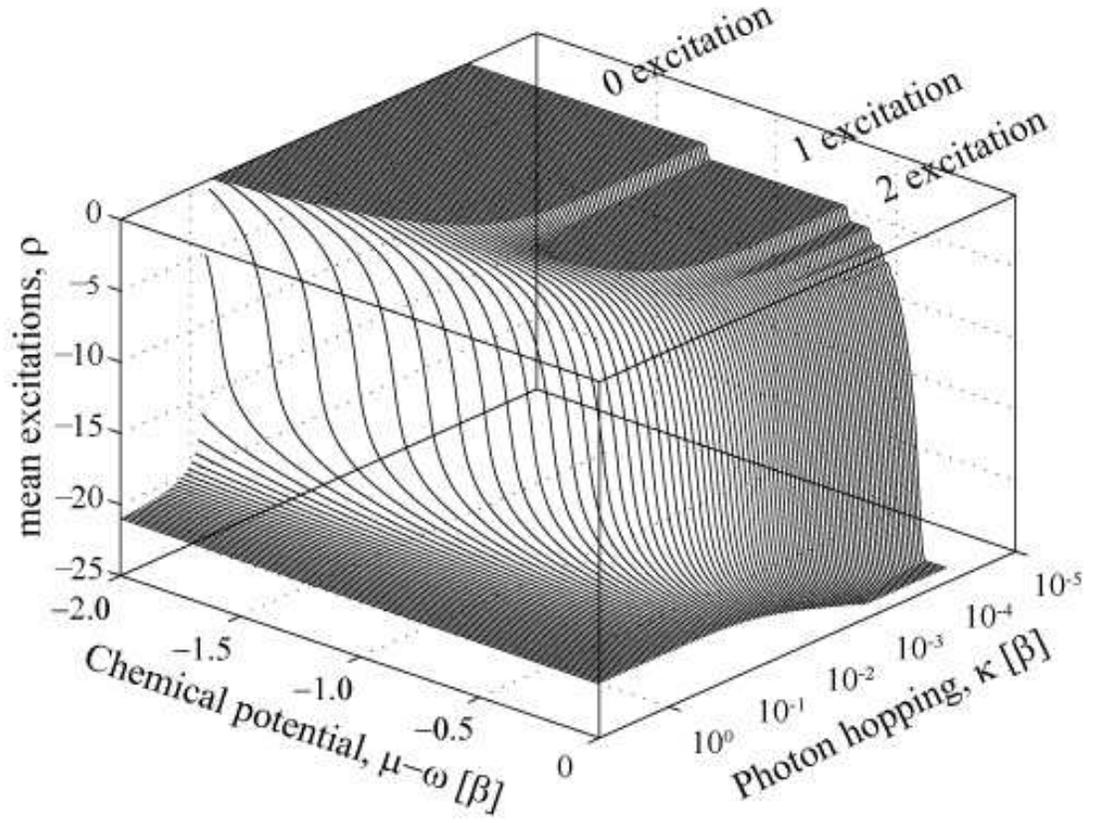}
\caption{\label{fig:dEgdmu} Plateaus with constant density, $\rho$, indicating
regions with a definite number state excitations, as a function of
the chemical potential, $\mu$, and photon hopping frequency, $\kappa$, for detuning, $\Delta = 0$.  The first three plateaus are
indicated, and the ground state configurations correspond to
$|g,0\rangle$, $|-,1\rangle$ and $|-,2\rangle$ for $0$, $1$, and $2$
excitations (photons in this case) respectively, with the plateaus shrinking in size with
increasing excitation number. Regions with varying $\rho$ have
coherent states as the ground state configuration.}
\end{figure}

To confirm the number of photons in each Mott lobe, we show in
Fig.~\ref{fig:dEgdmu} the average number of excitations per site in
the grand-canonical ensemble, $\rho = -\partial
E_{g}(\psi=\psi_{min})/\partial\mu$, as a function of $\kappa$ and
$\mu$ for $\Delta=0$. These plataues indicate regions of constant
density and incompresibility, both characteristic of the gapped Mott
insulator phase.

\section{Potential implementations}

Until now, we have deliberately avoided specific implementations, as
the above analysis should be applicable to all atom-photon systems
with strong coupling. In this section we specifically consider some
candidate implementations that may be realisable with present-day or
near-term technology.

One candidate system would be a photonic bandgap structure in
diamond, with individual negatively-charged nitrogen-vacancy (NV)
centres placed in each cavity. Micromachining of diamond is
presently being explored for similar purposes, see
\cite{bib:OliveroAdvMat2005,bib:BaldwinJVSTB2006}, and modelling has
shown that suitable cavities are in principle fabricatable
\cite{bib:THOptExp2006}. The application of diamond for quantum
computing and quantum optics applications has recently been reviewed
\cite{bib:GreentreeJPCM2006}. For NV diamond, the two-state
transition of interest is at
$637~\mathrm{nm}\sim3\times10^{15}~\mathrm{Hz}$. Calculations
presented in Ref.~\onlinecite{bib:GreentreePRA2006} showed that
$\lambda^{3}$ single-mode photonic bandgap cavities would have
$\beta\sim10^{10}\mathrm{Hz}$. Assuming that photon hopping limits
the cavity $Q$, we can approximate tunneling frequency by
$\kappa=\omega/Q$, so for $\kappa\leq\beta$ would require
$Q\geq10^{5}$. This is demanding, but we note that $Q\sim10^{7}$ has
been achieved in silicon-on-silica photonic-bandgap cavities
\cite{bib:SongNatMat2005}, although we stress that complete modeling
is necessary to precisely determine the required geometry. If the
diamond substrate is ultra-purity Type IIa (low nitrogen) diamond,
individual NV centres can be implanted using single-ion implantation
techniques \cite{bib:GreentreeJPCM2006,bib:JamiesonAPL2005}. Finally
we note that the resonance frequencies for the photonic bandgap
cavities will be extremely difficult to tune post-creation, however
the Stark effect can be used to tune the NV centres as required
\cite{bib:Tamarat} to allow an exploration (either statically or
dynamically) of the phase space shown in Fig.~\ref{fig:Fig3}.

Another candidate system where such effects may be observable is the
microwave strip line resonator for circuit QED recently demonstrated
by Walraff \textit{et al.} \cite{bib:WalraffNature2004}. In such
systems, the effective mode volume of the cavity can be much less
than $\lambda^{3}$, with $10^{-6}\lambda^{3}$ demonstrated in
Ref.~\onlinecite{bib:WalraffNature2004}, with the rate of loss of
photons out of the cavity (analogous to the $\kappa$ described in
this paper) being $\sim10^{3}~\mathrm{Hz}$. When such values are
combined with their atom-photon couplings, which were of order
$\beta\sim10^{7}~\mathrm{Hz}$, then the realised cavities would seem
to be well-suited to observing the effects we have described. One
advantage of the strip lines over the photonic bandgap case,
however, would be that generating structures with controlled (and
perhaps varying) numbers of nearest neighbours would be much easier.

The above two examples do not constitute the limit of possible
implementations. It may be possible to achieve these phase
transitions using Rydberg atoms in high-Q superconducting cavities.
Such systems have shown impressive results including generation of
one and two-photon states in the cavity \cite{bib:WeidingerPRL1999}.
An alternative architecture would be an array of microcavities
containing single atoms, with the cavities connected by optical
fibres \cite{bib:TrupkeAPL2005}. Again, this system would appear to
offer the desirable properties of high atom-photon coupling, with
relatively low hopping terms, and classical routing would permit
interesting lattice structures. Optical quantum dots, for example
InAs structural quantum dots in a photonic bandgap lattice, may also
afford interesting possibilities for the realisation of these (or
similar) quantum phase transitions.

\section{Conclusions}

We have shown how a Mott insulator to superfluid transition can be
realized with atom-dressed photons in a cavity superlattice.
Calculations for zero temperature give a phase diagram analogous to,
but fundamentally different from, the Bose-Hubbard model.

If one could Q-switch the cavities, for example by bringing each
cavity into close proximity to a near field probe, it may be
possible to out-couple all of the cavities simultaneously. This
would generate a self-ordered, two-dimensional array of single
photon emitters, which could be extremely beneficial for quantum
information applications. Because there is significant flexibility
in constructing exotic lattice geometries, one may be able to
engineer an extended Bose-Hubbard system suitable for topological
quantum computing \cite{bib:FreedmanTOPO} where fault-tolerance is a
natural consequence. Alternatively, simulation of quantum systems by
regime replication (where the Hamiltonian of the simulator system is
transformed to be equivalent to the Hamiltonian of some unknown
system) is an attractive application.

Similar though modified dynamics found in other Hubbard-like systems
should also be realizable in the atom-dressed photon superlattice,
such as a glass phase \cite{Fisher}, stripes and other symmetry
breaking phases \cite{Kosterlitz-Thouless}, super-solid behavior
\cite{scarola}, etc.  More importantly, since the excitation
spectrum of a fluctuation is gapped in the Mott-insulator region,
fluctuations due to temperature are exponentially suppressed. Thus,
one will be able to observe a phase that is formally not a Mott
insulator, but experimentally has very similar features
\cite{Fisher,Oosten2}. Disorder-induced destruction of the Mott
state \cite{Fisher} is also suppressed by this excitation-hole gap,
and system tunability as via the Stark shift may allow for manual
correction of these irregularities. Because of the Mott phase's
robustness, devices based on this effect at non-zero temperature
should be possible in this system. We also note that signatures of
quantum many-body phenomena should appear in 1D or finite arrays of
cavities.

\textit{Note Added:} Whilst we were preparing our manuscript, we
became aware of two related works, one by Hartmann \textit{et al.}
\cite{bib:HartmannQP2006}, and one by Angelakis \textit{et al.}
\cite{bib:AngelakisQP2006}, treating arrays of coupled quantum
cavities. The former \cite{bib:HartmannQP2006} considers a
four-state scheme for realising the required quantum nonlinearity
with time domain analysis. The latter \cite{bib:AngelakisQP2006}
performs a similar analysis explicitly considering two-state systems
per cavity in a 1-D arrangement to realise an XY chain. Neither
paper demonstrates Mott lobes as done here, nor did they perform a
mean field analysis. As such, both works are qualitatively different
from the present one, but demonstrate that photonic systems are ripe
for the exploration of condensed matter physical effects.

\section{Methods}

The eigenvalues and eigenvectors of the Hamiltonian of
Eq.~\ref{eq:JCHam} are
\begin{align}
|\pm,n\rangle & =
\frac{\left[-\frac{\Delta}{2}\pm\chi(n)\right]|g,n\rangle +
\beta\sqrt{n}|e,n-1\rangle}{\sqrt{2\chi^{2}(n)\pm\chi(n)\Delta}}~\forall
n\geq1,
\end{align}
with eigenvalues
\begin{eqnarray}
E_{|\pm,n\rangle} & = & n\omega\pm\chi(n)-\Delta/2.
\end{eqnarray}
Because $n\geq1$, the above definition for the dressed states does
not extend to $n=0$, and so we must define the ground state for the
dressed state system as $|g,0\rangle$, with $E_{|g,0\rangle}=0$.

To calculate the ground state wavefunction of our extended
Jaynes-Cummings Hamiltonian we must calculate the ground-state
energy $E_{g}$ as a function of $\psi$ as follows. First: obtain the
matrix elements of $H_{i}^{MF}$ in the onsite basis
$\{|g,0\rangle,\,|e,0\rangle,\,|g,0\rangle,\,|e,1\rangle,\,\ldots\}$
truncated at a finite value of $n$ states, $n_{max}$. Second:
diagonalize this matrix (in the truncated basis) and identify the
lowest eigenvalue, $E_{g}$. Third: increase $n_{max}$ until $E_{g}$
converges to its value at $n_{max}=\infty$. Finally: minimize
$E_{g}$ with respect to $\psi$ for different values of $\kappa$,
$\omega$, and $\Delta$ to obtain the phase diagram. At $T=0$ this
approach is operationally equivalent \cite{Sheshadri} to the
Gutzwiller Ansatz (GA) variational wave function technique
\cite{Rokshar,Krauth-GA} often used, but can be extended to finite
temperatures.

\begin{widetext}
For clarity, we will write out the first few terms of the mean-field
Hamiltonian in the onsite basis, $|g,0\rangle$, $|e,0\rangle$,
$|g,1\rangle$, etc.
\begin{eqnarray}
\mathcal{H}^{MF} = \left[
                     \begin{array}{ccccc}
                       \begin{array}{|c|} \hline
                       0 \\ \hline
                       \end{array} & \begin{array}{cc}
                              0 & -z \kappa \psi
                            \end{array}
                        &   &   &   \\
                       \begin{array}{c}
                         0 \\
                         -z \kappa \psi
                       \end{array}
                        & \begin{array}{|cc|} \hline
                             \epsilon - \mu & \beta \\
                             \beta & \omega - \mu \\ \hline
                           \end{array}
                         & \begin{array}{cc}
                              -z \kappa \psi & 0 \\
                              0 & -\sqrt{2} z \kappa \psi
                            \end{array}
                          &  &  \\
                        & \begin{array}{cc}
                              -z \kappa \psi & 0 \\
                              0 & -\sqrt{2} z \kappa \psi
                            \end{array} & \begin{array}{|cc|} \hline
                             \epsilon + \omega - 2\mu & \sqrt{2} \beta \\
                             \sqrt{2} \beta & 2\omega - 2\mu \\ \hline
                           \end{array} & \begin{array}{cc}
                              -z \sqrt{2} \kappa \psi & 0 \\
                              0 & -\sqrt{3} z \kappa \psi
                            \end{array} &  \\
                        &  & \begin{array}{cc}
                              -z \sqrt{2} \kappa \psi & 0 \\
                              0 & -\sqrt{3} z \kappa \psi
                            \end{array} & \begin{array}{|cc|} \hline
                             \epsilon + 2\omega - 3\mu & \sqrt{3} \beta \\
                             \sqrt{3} \beta & 3\omega - 2\mu \\ \hline
                           \end{array} & \ddots \\
                        &  &  & \ddots & \ddots \\
                     \end{array}
                   \right],
\end{eqnarray}
where the entries in the boxes correspond to the usual block
diagonal form of the Jaynes-Cummings Hamiltonian with the addition
of the superfluid order parameter, and the number of excitations in
each block increases by one as we move diagonally from block to
block from the ground state \(|g,0\rangle\).  The critical chemical
potentials can be immediately determined by equating the lower
eigenstates (negative dressed-state branch) from neighbouring
blocks, with $\kappa = 0$, which gives the result in
Eq.~\ref{eq:mucrit}.
\end{widetext}

\section{Acknowledgements}

The authors would like to acknowledge useful discussions with P.
Eastham, M. Friesen, D. Jamieson, R. Joynt, R. Kalish, P.
Littlewood, A. Martin, G. Milburn, J. Salzman, and H. Wiseman. CT is
funded by a USA National Science Foundation Math and Physical
Sciences Distinguished International Postdoctoral Research
Fellowship. This work was supported by the Australian Research
Council, the Australian Government and by the US National Security
Agency (NSA), Advanced Research and Development Activity (ARDA) and
the Army Research Office (ARO) under Contracts Nos. W911NF-04-1-0290
and W911NF-05-1-0284.

\section{Competing financial interests}

The authors declare that they have no competing financial interests.


\clearpage
\newpage
\onecolumngrid

\end{document}